\documentclass[11pt,oneside]{article}
\usepackage{graphicx}
\begin{document}
 
\title{{\bf A COMPILED CATALOGUE OF SPECTROSCOPICALLY DETERMINED ELEMENTAL ABUNDANCES
FOR STARS WITH ACCURATE PARALLAXES. MAGNESIUM}}
\author{{\bf T.V. Borkova, V.A. Marsakov}\\
Institute of Physics, Rostov State University,\\
194, Stachki street, Rostov-on-Don, Russia, 344090\\
e-mail: borkova@ip.rsu.ru, marsakov@ip.rsu.ru}
\date{accepted \ 2005, Astr. Rep., v.49, No 5}
\maketitle

\begin {abstract}
We present a compiled catalogue of effective temperatures, surface gravities,
iron and magnesium abundances, distances, velocity components, and
orbital elements for stars in the solar neighborhood. The atmospheric parameters
and iron abundances are averages of published values derived from model
synthetic spectra for a total of about 2000~values in 80~publications.
Our relative magnesium abundances were found from 1412~values in 
31~publications for 876~dwarfs and subgiants using a three-step iteration
averaging procedure, with weights assigned to each source of data as well as
to each individual determination and taking into account systematic deviations of 
each scale relative to the reduced mean scale. The estimated assumed completeness 
for data sources containing more than five stars, up to late December 2003,
exceeds 90\,\%. For the vast majority of stars in the catalogue, the 
spatial-velocity components were derived from modern high-precision 
astrometric observations, and their Galactic orbit elements were computed 
using a three-component model of the Galaxy, consisting of a disk, a bulge, 
and a massive extended halo.

\end {abstract}

\section*{Introduction}

Detailed study of the abundances of several elements in stars of 
various ages can be used to trace the Galaxy's chemical evolution 
and identify the sources in which various elements are synthesized, 
enabling progress in our understanding of the history of star 
formation and of the origin of our Galaxy's multi-component 
structure. According to current ideas, at least four subsystems can 
be identified in the Galaxy: the thin disk, thick disk, proto-disk 
halo, and accreted halo. It is believed that the first three 
subsystems formed from the same collapsing proto-galactic cloud, 
whereas the last subsystem was formed of isolated proto-galactic 
fragments and the debris of dwarf satellite galaxies that were 
disrupted and captured by the Galaxy at various stages of its 
evolution. (For more detail on the presence of accreted-halo stars 
in the solar vicinity, see Borkova, Marsakov~(2004) and references 
therein.) Since the vast majority of stars in the solar neighborhood 
are members of the thin-disk and thick-disk subsystems, the chemical 
compositions of these subsystems are best known. One striking result 
is that the thin-disk stars clearly exhibit a decrease in their 
relative abundances of $\alpha$\,-process elements, [$\alpha$/Fe], 
with increasing metallicity, [Fe/H ]. This means that the rate of 
iron enrichment of the interstellar medium during the thin-disk 
formation stage was greater than the rate of enrichment in 
$\alpha$\,-process elements. In addition, there is a strong difference 
between the thick-disk and thin-disk subsystems in their relative
abundances of $\alpha$\,-process elements,testifying that the
transition between the two disks was discrete (see, for instance,
Gratton et al.,~(1996),~(2000), Fuhrmann~(1998),~(2000)). Another 
interesting result is the absence of appreciable differences in the 
[$\alpha$\,/Fe] ratios for proto-disk halo and thick-disk stars. 
This can be interpreted as evidence that these two systems were
formed during a time so short that close binaries with component 
masses of 4--8\,M$_{\odot}$ did not have time to evolve, explode as 
type Ia supernovae, and enrich the interstellar medium with 
iron-peak elements. The large scatter in the relative abundances of 
$\alpha$\,-process elements in the halo is interpreted as evidence 
that mixing in the interstellar medium was weak during the formation 
of the halo subsystem (see, for instance, Fuhrmann~(2002)). The 
closeness of the ages of the stars in these two subsystems has even 
led some authors to doubt the existence of two separate subsystems,
and to suggest that the thick disk and proto-disk halo should be 
combined into a single subsystem (Fuhrmann~(2006), Gratton et al.~(2003). 
Note, however, that these conclusions are based on very limited samples,
with possible selection effects due to the extreme kinematic criteria 
used to isolate stars belonging to each of the subsystems. In fact,
a more careful study of the relative abundances of $\alpha$\,-process 
elements for thick-disk stars revealed a slight decrease of 
[$\alpha$/Fe] with increasing metallicity, considered to be evidence for 
the onset of enrichment of the interstellar medium with iron by type 
Ia supernovae Prochaska et al.~(2000). This means that the formation of the
thick disk began more than $\approx$\, 1 billion years after the first 
burst of star formation in the collapsing proto-Galactic cloud.
It was also found that a large number of stars with ratios 
[$\alpha$/Fe]$\leq$0.25 are observed in the range [Fe/H ]$\approx-1.0\dots-1.5$,
whichismore characteristic of thin-disk stars (see, for example, Nissen
\& Schuster~(1997)). In addition,most stars with unusually low 
abundances of $\alpha$\,-process elements have retrograde orbits and so 
were probably captured from dwarf satellite galaxies. This hypothesis 
is supported by the results of Shetrone et al.~(2003) and Tolstou et 
al.~(2003), who showed that stars with low [$\alpha$/Fe] values are actually
observed in dwarf spheroidal galaxies.

As we noted, the results cited above were obtained from fairly limited 
samples, which were often influenced by significant kinematic selection 
effects, and so require additional verification. The abundances of a 
number of elements have been published for numerous stars in the solar 
neighborhood. These were derived by different groups and have different 
accuracies. The best-studied $\alpha$\,-process element is magnesium:
unevolved F-–G stars possess optical lines of this element with various 
intensities produced by atoms in various excited states. Accordingly, the 
aim of our study is to compile a uniform master catalog of relative 
magnesium abundances for solar-vicinity stars reduced to a unified scale 
using essentially all published data,along with computed spatial-velocity
components and Galactic orbital elements based on accurate trigonometric 
parallaxes, proper motions, and radial velocities. This catalog will make 
it possible to extract samples in order to study various subsystems, not 
only with considerably larger volumes, but also largely free of artificial 
selection effects. Later, we will use these data together with theoretical 
evolutionary tracks to determine the ages of stars that have begun to leave 
the main sequence, taking into account the abundances of their 
$\alpha$\,-process elements. The high-accuracy kinematic data of our compiled
catalog will make it possible to reliably assign stars to particular 
subsystems of the Galaxy, and enable comprehensive statistical analyses of 
the stellar populations' chemical, physical, spatial, and kinematic 
properties. At the same time, homogeneous data on the relative magnesium 
abundances of stars currently in the solar vicinity but born at various 
distances from the Galactic center can be used to reconstruct the 
star-formation history and the evolution of the interstellar medium's 
chemical composition during the early stages of the formation of our Galaxy.

\section*{PREPARATION OF THE COMPILED CATALOGUE}

The various published abundances of an element for a given star often 
differ quite appreciably, even when the spectra reduced by different 
authors are of similarly high quality. Extraction of the needed 
information from the spectra is based solely on current ideas 
concerning the structure of stellar atmospheres, and these ideas 
develop via the usual iterative process. It is therefore not 
surprising that different authors have preferred to use somewhat 
different theoretical model stellar atmospheres and develop 
different techniques for reducing and theoretically analyzing spectra.
In several leading groups, analyses of spectral-line formation 
have been carried out without assuming local thermodynamic 
equilibrium. In fact, it is always necessary to analyze non-LTE 
effects for any abundance determinations. However, the assumption of 
LTE is justified for the atmospheres of dwarfs due to their high 
densities of particles, whose collisions bring the atomic-level 
populations into equilibrium. In particular, detailed computations 
have demonstrated that non-LTE corrections for magnesium in F-–G 
dwarfs and subgiants are small, within 0.1\,dex,and thus are smaller 
than the internal uncertainties in the Mg abundances Shimanskaya et 
al.~(2000). To avoid uncertainties due to non-LTE effects for lines
of iron, we always used abundances derived solely from Fe\,II lines.
(Note that, in his determinations of relative stellar magnesium 
abundances assuming LTE, Fuhrmann~(1998),~(2000) used lines of neutral 
elements for which the non-LTE effects are approximately the same,
and his resulting [Mg\,I/Fe\,I] ratios are nearly unbiased.)


\begin{table*}
\centering
\caption{The weights, $p_j$, and systematic deviations, $\Delta \textrm {[Mg/Fe]}_j$,
derived for each of the lists in two metallicity ranges} 
\tabcolsep0.6mm
\begin{tabular}{|l|r|c|c|c|c|c|c|}
\hline\hline
\multicolumn{1}{|c|}{Data sourse} & \multicolumn{1}{c|}{\parbox{1.5cm}{Total number of stars}} & 
\multicolumn{3}{c|}{\bf {[Fe/H]$\leq$-1.0}}& \multicolumn{3}{c|}{\bf {[Fe/H]$>$-1.0}}\\
\cline{3-8}
& & \bf{$\Delta$[Mg/Fe]$_j$} & \bf{p$_j$} & {\bf{N}} & \bf{$\Delta$[Mg/Fe]$_j$} & 
       \bf{p$_j$} & {\bf{N}}\\
\hline
Edvardsson et al. (1993) & 178 & --     & --   & -- & -0.045 & 1.00 & 167 \\
Nissen \& Schuster (1997)& 30  & +0.001 & 1.00 &  7 & +0.046 & 0.60 & 22  \\
Mashonkina et al. (2003) & 61  & -0.056 & 0.64 & 27 & -0.026 & 0.63 & 32  \\
Øèìàíñêàÿ (2000)         & 15  & +0.034 & 1.00 & 13 & -0.069 & 0.63 & 2   \\
Fuhrmann (1998, 2000)    & 111 & -0.040 & 0.52 &  6 & +0.006 & 0.81 & 76  \\
Fuhrmann (1995)          & 52  & +0.053 & 0.60 & 20 & +0.053 & 0.63 & 22  \\
Chen et al. (2000, 2003) & 93  & --     & --   & -- & -0.011 & 0.64 & 59  \\
Mishenina (2001, 2004)   & 209 & +0.070 & 0.67 & 21 & +0.004 & 0.69 & 77  \\
Clementini et al. (1999) & 16  & -0.0035& 0.61 & 7  & +0.041 & 0.49 & 7   \\
Carretta et al. (2000)   & 9   & --     & --   & -- & +0.008 & 0.70 & 5   \\
Nissen (1994)            & 8   & -0.030 & 0.96 & 7  & --     & --   & --  \\
Gehren (1998, 2000, 2004)& 51  & -0.019 & 0.57 & 16 & +0.011 & 0.46 & 21  \\
Magain (1989)            & 20  & -0.111 & 0.43 & 20 & --     & --   & --  \\
Jehin et al. (1999)      & 20  & --     & --   & -- & +0.017 & 1.00 & 20  \\
Stephens (1999, 2002)    & 55  & -0.082 & 0.28 & 8  & --     & --   & --  \\
Carney et al. (1997)     & 7   & -0.050 & 0.63 & 7  & --     & --   & --  \\
Idiart et. al. (1999)    & 217 & +0.129 & 0.40 & 40 & +0.052 & 0.80 & 176 \\
Procheska et al. (2000)  & 10  & --     & --   & -- & +0.040 & 1.00 & 8   \\
Bensby et al. (2003)     & 65  & --     & --   & -- & -0.002 & 0.66 & 27  \\
Åðìàêîâ (2002)           & 20  & +0.098 & 0.33 & 11 & -0.080 & 0.76 & 4   \\
Rayn et al. (1991, 1996) & 27  & -0.030 & 0.33 & 9  & --     & --   & --  \\
Gratton et al. (2003)    & 142 & -0.077 & 0.67 & 41 & -0.076 & 0.77 & 53  \\
Reddy et al. (2003)      & 176 & --     & --   & -- & +0.046 & 0.90 &  30 \\
\hline
\end{tabular}
\begin{list}{}{}
  \item[Note.] The "N" columns give thenumber of a list's stars that are
  also present in other sources
\end{list}
\end{table*}

These factors explain the reasons for the deviations among the data 
obtained and analyzed by different authors. We also cannot rule out the 
possibility of systematic differences between the magnesium abundances 
presented in different papers. If several abundance values are 
available for the same star, they can simply be averaged. However, when 
an abundance is presented in only one paper, the possibility of 
systematic differences must be considered. We collected all available 
lists (with $\geq$\,5\,stars) of relative magnesium abundance estimates,
[Mg/Fe], for field stars from high-resolution spectra with high 
signal-to-noise ratios published after 1989. We estimate the 
completeness of the abundances published for solarvicinity stars up 
through December 2003 to be better than 90\,\%. The largest of the 
source catalogs contain only about 200 stars (Table 1). Note that our 
aim is not to analyze the origin of discrepancies among the data,
but to compile a list of published spectroscopic relative magnesium 
abundances for field stars that is as complete as possible, and is 
reduced to a uniform scale. The raw material for this study were 
36~publications containing 1809~magnesium-abundance determinations 
for 1027~stars. Of these stars,we retained only 876~dwarf and subdwarf 
stars that lie below the solid line in the Hertzsprung–Russell diagram 
in Fig.\,1a. This line was plotted "by eye,"parallel to the zero-age 
main sequence, to exclude stars in more advanced evolutionary stages.
This automatically rejected five publications in which only giants 
had been studied.

\section*{Atmospheric Parameters and Iron Abundances}

Atmospheric Parameters and Iron Abundances We simply averaged the 
stellar effective temperatures and surface gravities from the cited 
papers, in which these parameters were determined using various methods.
For both parameters, we estimated the uncertainties of the averages 
based on the scatter of the individual values about the average for 
each star; i.\,e., from the agreement of the values obtained by the 
various authors. For this purpose,we calculated the deviations 
$dX_i=\langle X\rangle-X_i$ for stars for which these parameters were 
determined in several studies, where the index i refers to the 
individual measurements for a given star. Note that our uncertainty 
estimates are based on as many as 80~publications, since we also used 
here studies devoted to spectroscopic abundance determinations for 
other elements (see the numbers of stars in the histograms, Fig.\,2).
We then plotted the corresponding distributions and calculated the 
dispersions of Gaussian curves describing them, which were equal to 
$\varepsilon$\,(T$_{\textrm{eff}})=\pm$\,56\,K and $\pm$\,82\,K and 
$\varepsilon (\log g)= \pm$\,0.12 and $\pm$\,0.24 for stars with
[Fe/H] values above and below $-1.0$, respectively. The good 
agreement of the observed distributions with the Gaussian curves 
shows that the deviations are random, so that their dispersions 
should reflect the actual uncertainties in the parameters. Our 
[Fe/H] values for each star are also averages of metallicities
from the same papers,with uncertainties estimated as 
$\varepsilon$\,([Fe/H])\,=\,$\pm$\,0.07\,dex and $\pm$\,0.13\,dex,
respectively, for metal-rich and metal-poor stars. All these 
estimates are close to the lower limits of the uncertainties for 
these parameters claimed by the authors. These distributions and 
the approximating normal curves are displayed in Fig.\,2. We found 
it necessary to differentiate between the two metallicity groups 
because the uncertainties in all the parameters are considerably 
larger for the metal-poor stars.

\section*{Iterative Procedure Used to Calculate the Relative Magnesium
Abundances}

To get an idea of the number of abundance determinations for the 
individual stars, see the histogram in Fig.\,1\,b, which presents the
distribution of the number of sources for the stars. This histogram 
shows that the number of stars decreases exponentially with increasing 
number of magnesium-abundance determinations, so that only single 
determinations are available for more than half the stars in the 
sample. To derive reliable abundances, we applied a somewhat modified 
version of the three-step iterative technique for compiling data 
presented by Castro et al.~(1997).

Figure~3 presents several diagrams that compare the [Mg/Fe] ratios 
for the lists having the largest numbers of stars in common. We can 
see that the values of all the authors are fairly well correlated,
with all the correlation coeficients being $r \geq 0.8$. However,
small (within the uncertainties)systematic deviations are present 
in the data. Not only are the results of the LTE and non-LTE 
approximations different, but there exist discrepancies within each 
of these approximations. In particular,the two diagrams in the upper 
panel of Fig.\,3 demonstrate the scatter of the estimates obtained 
using a single approximation. The two diagrams in the middle panel,
which compare the abundances derived using different approximations,
show only scatter without a considerable systematic offset. On the 
other hand, appreciable systematic differences can appear not only 
between the values obtained using different approximations, but also 
between the values obtained by different authors using the same 
approach (Figs.\,3e and 3f). If there are no regular systematic 
differences related to a particular approach, we can simply average 
all these data for each star.

After averaging (calculating the individual preliminary mean values,
$\langle$\,[Mg/Fe]$_i\rangle$), we can check whether the systematic 
differences and scatter of the [Mg/Fe]$_i$ values for different 
authors depend on metallicity. Figure\,4 shows the relations between
the deviations $\delta$\,[Mg/Fe]$_i=\langle$\,[Mg/Fe]$_i\rangle$\,-\,[Mg/Fe]$_i$
and [Fe/H]$_i$ for some of our lists. The subscript {\it i} here 
refers to an individual abundance determination for a star. The 
diagrams demonstrate that the scatter of the deviations and the 
systematic offset of $\delta$\,[Mg/Fe]$_i$ relative to the mean 
(zero) vary from list to list and also depend on metallicity.
Sometimes the scatter and systematic offset remain practically
the same over the whole metallicity range (Figs.\,4a and 4b). Weak 
trends with various signs are common. The $\delta$\,[Mg/Fe]$_i$ 
deviations sometimes move upwards (Figs.\,4c and 4d)and sometimes 
downwards (Figs.\,4e and 4f) with increasing metallicity in the
diagram. To take these small but systematic trends into account,
we divided each list into two metallicity magnesium-ranges at 
[Fe/H]\,$=-1.0$ and calculated the mean deviations for these 
ranges; i.\,e., the systematic deviations 
$\Delta$\,[Mg/Fe]$_i =\langle\delta$\,[Mg/Fe]$_j\rangle$ (see the 
dotted lines in the diagrams). We then corrected all the individual
[Mg/Fe]$_i$ values for these biases. As noted above, these 
corrections leave the magnesium abundances for stars present in 
several lists virtually unchanged. However, if a star's magnesium 
abundance was determined in a single study only, the correction
will strongly affect the final magnesium abundance. Indirect 
confirmation that it is appropriate to correct for these systematic 
differences is provided by the fact that the scatter of the 
data points in the final [Fe/H]–-[Mg/Fe] diagram plotted for the 
sample stars became considerably smaller after applying the 
corrections, and characteristic features observed in the data 
for individual sources became more prominent (compare Fig.\,10 below 
and Fig.\,6 in Mashonkina et al. (2003)).

The next step after correcting for systematic biases was to 
determine weights for the data sources in the lists and 
calculate new weighted means. Figures~3 and 4 show that the 
scatters in the diagrams —- i.\,e., the [Mg/Fe] uncertainties 
in the lists —- are not all the same. Accordingly, each list 
was assigned a weight that was inversely proportional to the 
corresponding dispersion for the deviations in each of the 
metallicity ranges. The lowest scatter for the higher 
metallicities was found for the lists of Prochaska et al.~(2000), 
Edvardsson et al.~(1993), Jehin et al.~(1999), and they were 
assigned unit weights. At lower metallicities, the lowest
scatter was shown by the lists of Nissen \& Schuster~(1997),
Mashonkina et al.~(2003). The lowest weights assigned to some 
of the lists were $\approx$\,0.3. The weights and biases for 
each list in each of the two metallicity ranges are collected 
in Table\,1. We then calculated a new weighted mean magnesium 
abundance for each star taking into account the biases and
weights assigned to the lists.

Let us briefly discuss one important aspect of compiling the 
data. In practice, small catalogs can have few stars in common 
even with the largest original catalogs. Thus, to analyze the 
agreement between their values and the values of other authors
and assign them correct weights,we combined the small catalogs 
into larger lists, each containing a series of studies by the 
same group in which the same techniques were used to derive the 
abundances. For lists containing several values for the same star,
we chose the more recent values. As a result, we obtained 24~lists 
including 31~primary data sources (Table\,2\footnote{Table~2 of 
this paper is presented only electronically and can be found at 
http://cdsweb.u-strasbg.fr/cats/J.AZh.htx. A description of this 
table is given in the Appendix.}). Further, all the lists were 
used to calculate preliminary mean magnesium abundances for each
star in the sample. (The only exception was Castro et al.~(1997): 
none of its nine stars were found in any other source, and all 
the data for these stars were entered into the final catalog 
without any changes.)

The next step was also a weight-assigning procedure, this time 
for individual values, [Mg/Fe]$_{ij}$, where the first subscript 
refers to the value and the second to the list. This procedure 
was intended to assign lower weights to initial values showing 
larger deviations. Clearly, such a procedure can work only if 
there are three or more values for the same star. This step 
makes use of the deviations from the calculated weighted means,
$\delta_{ij}$, for individual values (with their biases taken 
into account). When assigning the weights $p_{ij}$, we considered 
the mean absolute value, $\varepsilon_j$, of the deviations for 
all stars in the list containing the given value. The weights 
were calculated using the formula 
$p_{ij}=\sqrt{2\varepsilon^2_j/(\delta^2_{ij}+\varepsilon^2_j)}$.
We can see that an individual weight of unity was assigned to 
values whose deviations were equal to the mean deviation. Most 
stars in the list were given weights slightly greater than unity,
$(p_{ij})_{max}=\sqrt{2}$. However, the weights begin to decrease 
appreciably for $\delta_{ij}>\varepsilon_j$.

As a result, this procedure assigns the lowest weights to the 
least-reliable determinations and enables us to obtain final values
that are close to those given for most of the sources, with no 
single measurement rejected. Some idea of the internal accuracy of 
our final [Mg/Fe] values can be obtained from the dispersion of 
the distribution of all the deviations $\delta_{ij}$, equal to 
$\varepsilon$\,[Mg/Fe]\,$=0.05$ and 0.07\,dex for metal-rich and 
metal-poor stars, respectively (Fig.~5). The final weighted mean 
[Mg/Fe] values calculated for our sample of dwarfs and subgiants 
in the solar neighborhood are presented in Table (see the Appendix).

\section*{Distances, Proper Motions, and Radial Velocities}

We determined the distances to the stars using trigonometric 
parallaxes with uncertainties below 25\,\% from the catalogs 
Hipparcos~(1997), Kharchenko~(2001), Myers et al.~(2002) or, in 
their absence, we adopted the photometric distances from the
catalogs Carney et al.~(1994), Nordstrom et al.~(2004), derived 
using uvby$\beta$ photometric data. The uncertainty in photometric 
distances is usually claimed to be $\pm$\,13\,\% Nordsrom et al.~(2004).
Figure~6 shows the distribution of the distances of the sample 
stars from the Sun. The distribution's ascending branch lasts 
only to 5\,pc, after which a steep decline begins, testifying 
that most of the sample stars are close to the Sun. The photometric 
distances were used only for the most distant stars.

We took the proper motions from the catalogs Hipparcos~(1997), 
Kharchenko~(2001), Myers et al.~(2002), Beer et al.~(2000), 
Carney et al.~(1994), Bakos et al.~(2002) the order of the 
catalogues corresponds to priority of utilizations). For most 
stars in our sample, the typical uncertainties in their proper
motions are $\approx$\,1.8\,milliarseconds/year, corresponding 
to a mean uncertainty in the tangential velocity of 
$\approx\pm$\,0.8\,km/s.

We took the radial velocities from the catalogs Nidever et al.~(2002),
Barbier-Brossat \& Figon~(2000), Barbier-Brosat et al.~(2002), 
Nordstrom et al.~(2004)the order of the catalogues corresponds to 
priority of utilizations). If the required data were not present 
in any of these catalogs, we took them from other publications (see
references to Table in the Appendix). The characteristic 
uncertainties in the radial velocity measurements for the sample 
stars are $\approx \pm$\,0.6\,km/s (Myers et al~(2002).

\section*{Spatial Velocities and Galactic Orbital Elements}

Spatial Velocities and Galactic Orbital Elements We computed the 
$U$, $V$, and $W$ components of the total spatial velocity 
relative to the Sun for 850~stars with distances,proper motions,
and radial velocities ($U$ is directed towards the Galactic 
anticenter, $V$ in the direction of the Galactic rotation, and 
$W$ toward the North Galactic pole). The main contribution to
the uncertainties in the spatial velocities comes from the 
uncertainties in the distances, rather than the uncertainties 
in the tangential and radial velocities. For mean distance 
uncertainties of 15\,\% and the mean distance from the Sun of 
the sample stars, $\approx 60$\,pc, the mean uncertainty in 
the spatial velocity components is $\approx \pm 2$\,km/s.

We adopted a Galactocentric distance for the Sun of 8.5\,kpc,
a rotational velocity for the Galaxy at the solar Galactocentric 
distance of 220\,km/s, and a velocity for the Sun relative to 
the Local Standard of Rest of 
($U_{\odot}, V_{\odot}, W_{\odot})=(-11, 14, 7.5$)\,km/s 
Ratnatunga et al.~(1989). We used these values to calculate 
the components of the spatial velocities in cylindrical 
coordinates, ($\Theta, \Pi, W$), and the total residual 
velocities of the stars relative to the Local Standars of Rest,
$V_{\textrm{res}}$.

We calculated the Galactic orbital elements by modeling 30~orbits 
of each star around the Galactic center using the multi-component 
model for the Galaxy of Allen \& Santillan~(1991), which consists 
of a disk, bulge, and extended massive halo.

A list of the stars with all the parameters described in this paper 
is presented in Table~2 (see its description in the Appendix). The 
catalog itself is published only electronically, at the address 
http://cdsweb.u-strasbg.fr/cats/J.AZh.htx

\section*{ANALYSIS OF THE SAMPLE}

All the published catalogs can be tentatively subdivided into two 
groups based on the principle applied to select stars for 
spectroscopic determinations of the abundances of various elements.
The first group consists of various samples whose stars were 
selected based on a limiting apparent magnitude. Since the Sun is 
situated near the Galactic plane, such samples contain almost 
exclusively stars of the disk subsystems, with very few halo (i.\,e.,
low-metallicity) stars. Thus, a preliminary selection of 
low-metallicity stars using photometric data is needed for 
spectroscopic observations of objects in the halo subsystems. The
apparent magnitudes of the stars selected in this way are much 
fainter. This is clearly illustrated in the [Fe/H]-–V diagram for 
our sample stars (Fig.~7). Nearly all the stars with 
[Fe/H ]\,$\geq -1.0$ are brighter than V$\approx 9^m$, whereas 
most of the more metal-poor stars are fainter. The stars of both 
groups cover the sky rather uniformly, with an understandable 
concentration of metal-richer stars toward the Galactic plane
(Fig.~8). This observational selection effect should be born in 
mind when estimating the relative numbers of stars of various 
metallicities in the solar neighborhood. Figure~9 displays the 
[Fe/H] and [Mg/Fe] distributions for the resulting sample, while 
Fig.~10 shows the corresponding [Fe/H]-–[Mg/Fe] diagram.

We are preparing a cycle of papers in which we identify stars 
belonging to the thin-disk and thick-disk subsystems, as well 
as to the proto-disk and "accreted" halo subsystems, based on 
the elements of the Galactic orbits from the catalog. In these 
forthcoming papers, we will estimate the star-formation rates 
and efficiencies of mixing of the interstellar medium ing the 
early stages of the Galaxy's evolution based on an analysis of 
the scatter of the relative magnesium abundances for 
low-metallicity stars of the old Galactic subsystems.

We are also planning to present similarly unified abundance 
determinations for other $\alpha$\,-process elements in a future 
version of our catalog.

{\bf ACKNOWLEDGMENTS}:
The authors are grateful to L.I.Mashonkina for fruitful 
discussions of the results.

\newpage
\begin{flushright}
\emph{Appendix}
\end{flushright}

The description of Table~2 (this Table is being published only 
electronically, at the address \\
http://cdsweb.u-strasbg.fr/cats/J.AZh.htx):\\

\noindent (1) HD: No. in the HD catalogue $\langle$ III/135 $\rangle$\\
(2) DM: DM number $\langle$I/119$\rangle$, $\langle$I/122$\rangle$\\
(3) HIP: No. in the Hipparcos catalogue ($\langle$I/239$\rangle$)\\
(4) SAO: No. in the SAO catalogue $\langle$I/131$\rangle$\\
(5) OtherName: Other name\\
(6) Vmag:  Apparent $V$ magnitude in the Johnson system \\
(6) Radeg: $\alpha$ in degrees (ICRS, epoch J1991.25)\\
(7) Dedeg: $\delta$ in degrees (ICRS, epoch J1991.25)\\
(8) Plx:   Trigonometric or photometric parallax

\parindent=1cm \emph{Remark to column 8:}

\parindent=1cm Errors are given for trigonometric parallaxes only.\\
(9) pmRA:  Proper motion, $\mu_\alpha\cos\delta$ \\
(10) pmDE: Proper motion, $\mu_\delta$ \\
(11) e\_Plx:  Standard error of the trigonometric parallax\\
(12) e\_pmRA: Standard error of $\mu_\alpha\cos\delta$ \\
(13) e\_pmDE: Standard error of $\mu_\delta$ \\
(14) f(pi): Data source for the star's parallax and proper motion

\parindent=1cm \emph{Remark to column 14:}

\parindent=1cm The following notation is used for the references:
\parindent=1cm\hangindent=1cm [HIP]: The Hipparcos and Tycho 
           Catalogues, ESO, 1997

\parindent=1cm\hangindent=1cm  [KHR]: Kharchenko N.V., 2001, 
           All-sky Compiled Catalogue of 2.5 million  stars,  KFNT, 17, 
           409

\parindent=1cm\hangindent=1cm [SKY]: Myers J.R., Sande C.B., Miller A.C., 
           et al., 2002, Sky2000. Catalogue, version 4

\parindent=1cm\hangindent=1cm [BCY]: Beer T.C., Chiba M., Yoshi Y., 
           et al., 2000, Astron J., 119, 2866

\parindent=1cm\hangindent=1cm [CLL]: Carney B.W., Latham D.W., Laird J.B., 
           Aguilar L.A., 1994, Astron. J., 107, 2240

\parindent=1cm\hangindent=1cm [BSN]: Bakos G.A., Sahu K.C., Nemeth P., 
           2002, Astrophys. J. Suppl. Ser. JS, 141, 187.\\

\parindent=0cm\hangindent=0cm (15) RV:    Radial velocity\\
\parindent=0cm\hangindent=0cm(16) f(vr): Data source for the radial velocity (3)

\parindent=1cm \emph{Remark to column 16:}

\parindent=1cm The following notation is used for the references:

\parindent=1cm\hangindent=1cm [BBF]: Barbier-Brossat M. and Figon P., 
           Astron. and Astrophys. 142, 217, 2000

\parindent=1cm\hangindent=1cm [BG]: Barbier-Brossat M. and Gratton R.G., 
           Astron. and Astrophys., 384, 879, 2002

\parindent=1cm\hangindent=1cm[MLG]: Malaroda S., Levato H., Galliani S., 
           Stellar Radial Velocities 1991--1998

\parindent=1cm\hangindent=1cm [NMB]: Nidever D.L., Marcy G.W., 
           Butler R.P., et al. Astron. J. Suppl. Ser., 141, 503 (2002)

\parindent=1cm\hangindent=1cm [NMA]: Nordstrom B., Mayor M., Andersen J., 
           et al., Astron. and Astrophys., 418, 989 (2004)

\parindent=1cm\hangindent=1cm [MON]: Montes D., Lopez-Santiago J., 
           Galvez M.C., et al., Monthly NOtices Roy. Astron. Soc., 328, 
           45 (2001)

\parindent=1cm\hangindent=1cm [FUL]: Fulbright J.P., Astron. J., 123, 
           404 (2002)

\parindent=1cm\hangindent=1cm [WBS]: Wilhelm R., Beers T.C., 
           Sommer-Larsen J., et al., Astron. J., 117, 1229 (1999)

\parindent=1cm\hangindent=1cm [SWG]: Strassmeier K.G., Washuettl A., 
           Granzer T., et al., Astron. and Astrophys. Suppl. Ser., 142, 
           275 (2000)

\parindent=1cm\hangindent=1cm [LST]: Latham D.W., Stefanik R.P., 
           Torres G., et al., Astron. J., 124, 1144 (2002)

\parindent=1cm\hangindent=1cm [NS]: Nissen P.E., Schuster W.J. Astron. 
           and Astrophys., 326, 751 (1997)

\parindent=1cm\hangindent=1cm [GBR]: Grenier S., Baylac M.O., 
           Rolland L., et al., Astron. and Astrophys. Suppl. Ser., 137, 
           451 (1999)

\parindent=1cm [AFG]: Axer M., Fuhrmann K., Gehren T., Astron. and 
           Astrophys., 300, 751 (1995). \\
           
\parindent=0cm\hangindent=0cm (17) Teff:   Effective temperature \\
\parindent=0cm\hangindent=0cm (18) log\,g:  The star's surface gravity\\
\parindent=0cm\hangindent=0cm (19) [Fe/H]:  The star's iron abundance (4)\\
\parindent=0cm\hangindent=0cm (20) [Mg/Fe]: The star's magnesium-to-iron abundance

\parindent=1cm\hangindent=1cm \emph{Remark to columns 19, 20:}

\parindent=1cm\hangindent=1cm  Abundances of chemical elements are given in
           the logarithmic scale, expressed in solar units.

\parindent=0cm\hangindent=0cm (21) f(mg): Number of data sources for [Mg/Fe] \\
\parindent=0cm\hangindent=0cm (22) f(sum): The data sources' combined weight

\parindent=1cm\hangindent=1cm \emph{Remark to column 22:}

\parindent=1cm\hangindent=1cm  The combined weight includes the weight of the
data source (Table~1) and of the individual value (see the text).

\parindent=0cm\hangindent=0cm   (23) l:  Galactic longitude \\
(24) b:   Galactic latitude \\
(25) Mv:   Absolute magnitude\\
(26) rsun:  Distance from the Sun\\
(27--29) Xg, Yg, Zg:  Coordinates in the galactic coordinate system\\
(30) R:    Distance to the star from the Galaxy's center \\
(31-33) U, V, W: The stars space velocity components (see the text)\\
(34, 35) Vpi, Vtet: Components of the space velocity in the cylindrical
coordinate system\\
(36) Vpec: The stars total peculiar velocity with respect of the Local Standard
of Rest\\
(37--39) Rpe, Rap, e: The stars orbital perigalactic and apogalactic radius and
orbital eccentricity\\
(40) $\Psi$: Inclination of the star's galactic orbit\\
(41) Zmax: Maximal distance of the star's orbit from the galactic plane.\\

\newpage

\begin{figure*}
\centering
\includegraphics[angle=0,width=16cm,bbllx=80pt,bblly=10pt,bburx=650pt,bbury=770pt]{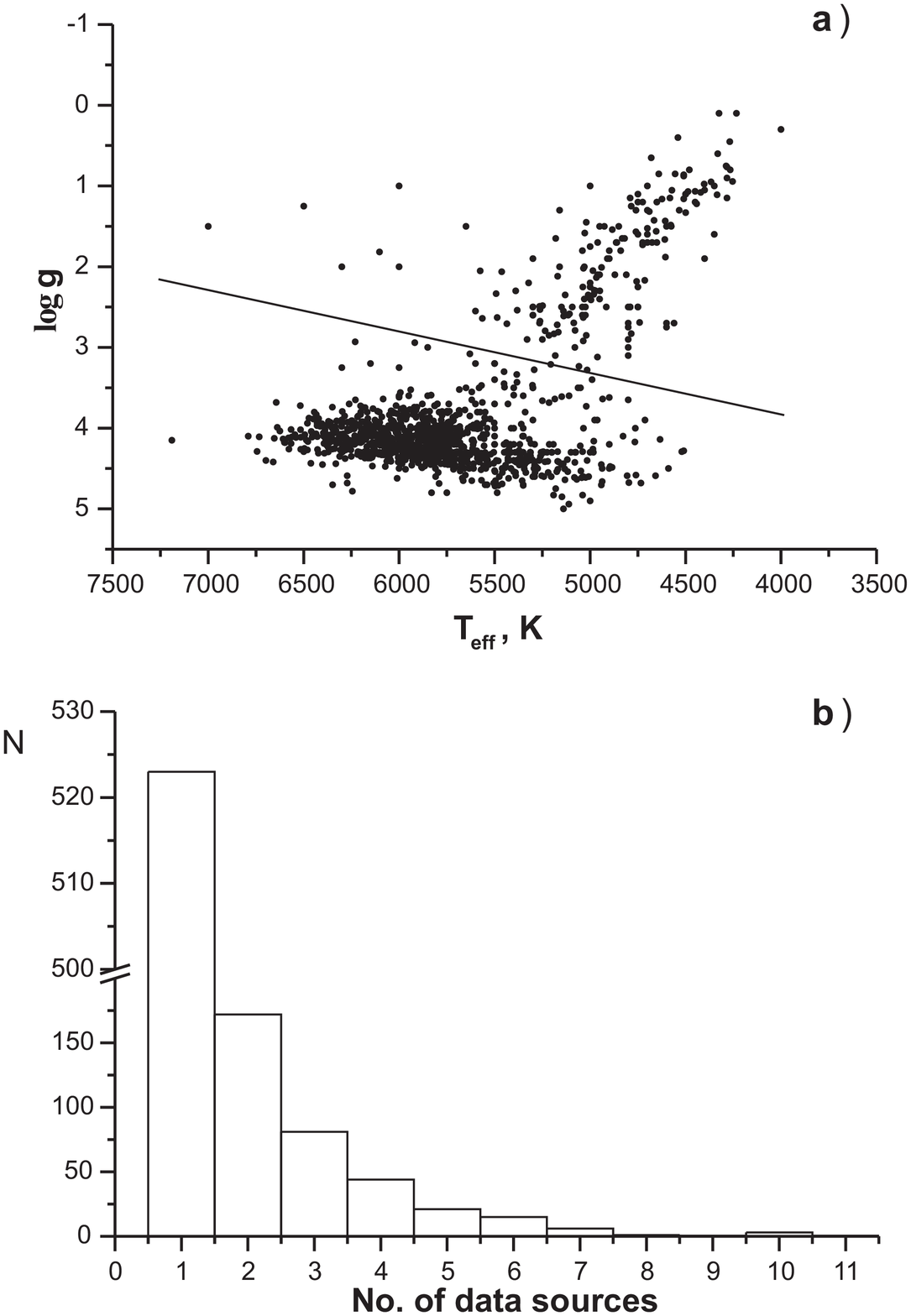}
\caption{The  $T_{\textrm{eff}}-\log g$ diagram for stars with magnesium
abundances found in the literature (a) and the stars' distribution over the
number of magnesium-abundance determinations available for them (b). Only
stars that lie below the sloping line in the diagram (a) enter our catalogue.
\hfill
}
\label{fig1}
\end{figure*}


\begin{figure*}
\centering
\includegraphics[width=15cm,angle=90,bbllx=100pt,bblly=100pt,bburx=700pt,bbury=480pt]{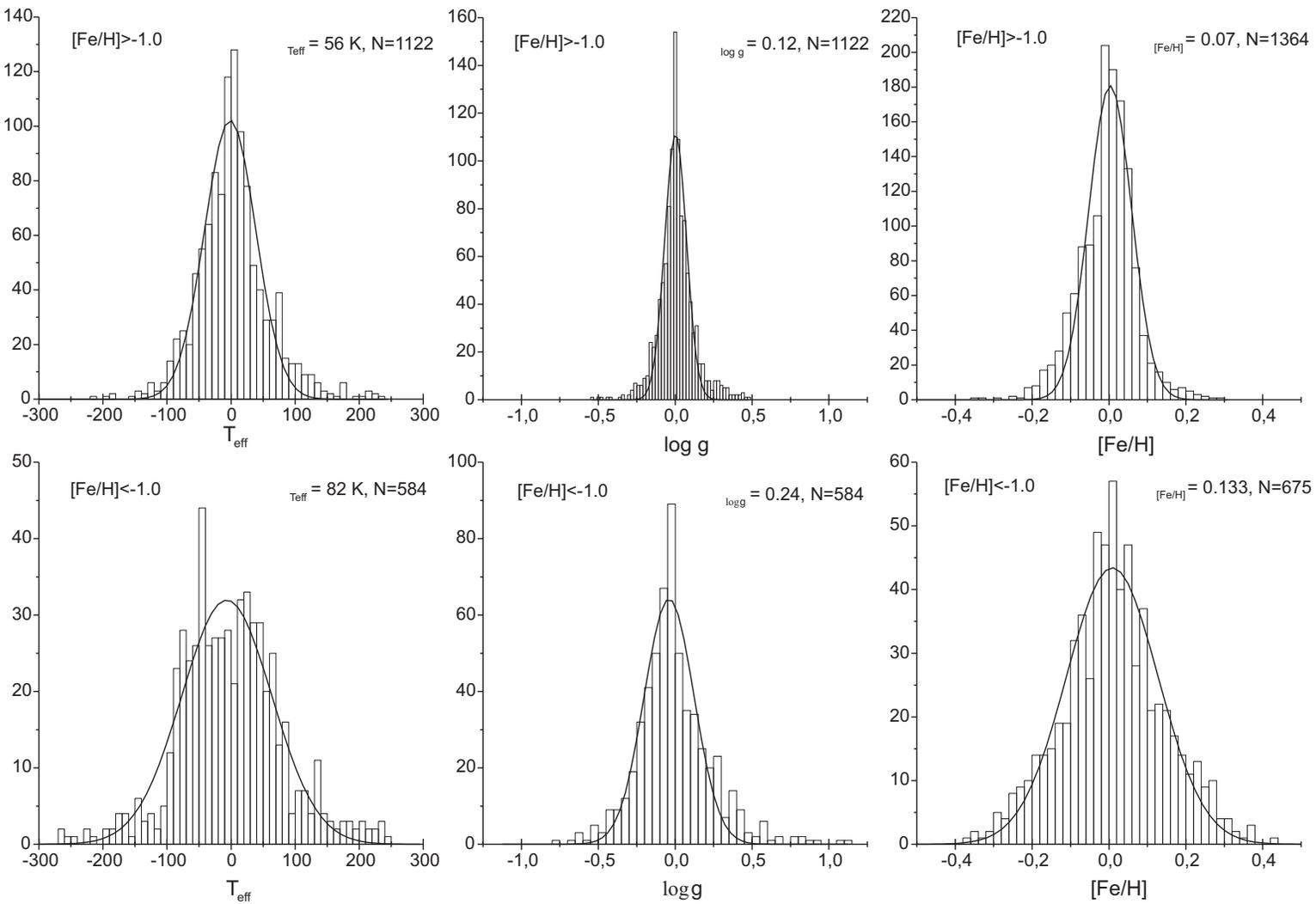}\vspace{30mm}
\caption{The distribution of deviations of individual determinations of
$T_{\textrm{eff}}$, $\log g$, and [Fe/H] from the computed mean values. The
upper row of the graphs is for the stars with $[\textrm{Fe}/\textrm{H}]>-0.1$,
and the lower row, for $[\textrm{Fe}/\textrm{H}] < -1.0$. The curves
approximate the distributions with a normal law. Dispersions and star numbers
are indicated for the histograms. \hfill}
\label{fig2}
\end{figure*}

\begin{figure*}
\centering
\includegraphics[angle=0,width=0.99\textwidth,clip]{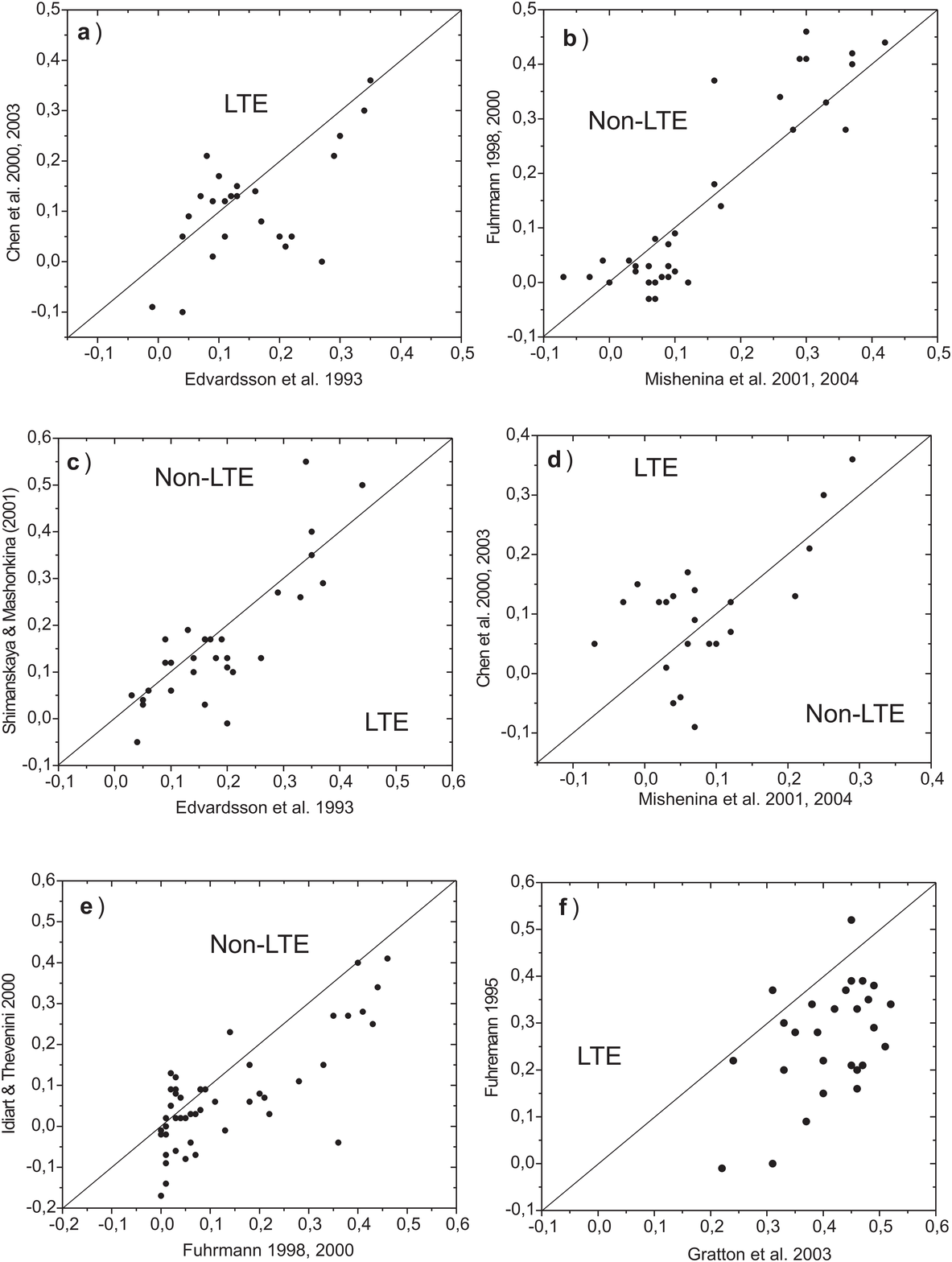}
\caption{Comparison of [Mg/Fe] values from several authors. The
approximation used by the corresponding authors for magnesium-abundance
determinations is indicated.
\hfill}
\label{fig3}
\end{figure*}

\begin{figure*}
\centering
\includegraphics[angle=0,width=0.99\textwidth,clip]{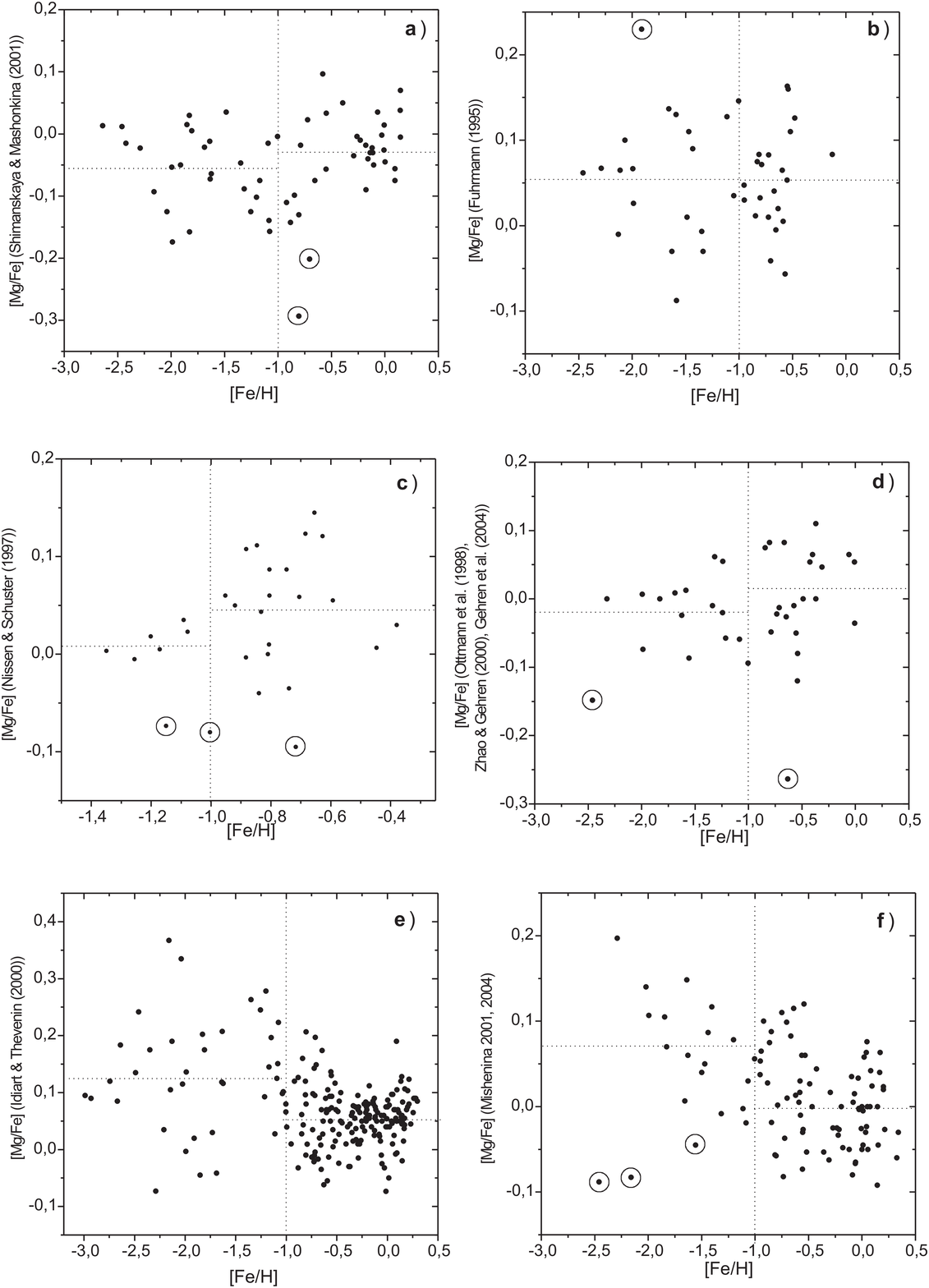}
\caption{The relation between metallicities and deviations of individual
magnesium-abundance determinations (from several sources) from those computed
as averages from all the authors, ($\delta$[Mg/Fe]). The values rejected using
the 3\,$\sigma$ criterion in the scatter
($\sigma_{\delta[{\textrm{Mg}}/{\textrm{Fe}}]}$) or deviation
($\Delta$[Mg/Fe]$_j$) computations are circled. The horizontal lines
are mean deviations for each of the ranges.
\hfill}
\label{fig4}
\end{figure*}

\begin{figure*}
\centering
\includegraphics[angle=0,width=0.99\textwidth,clip]{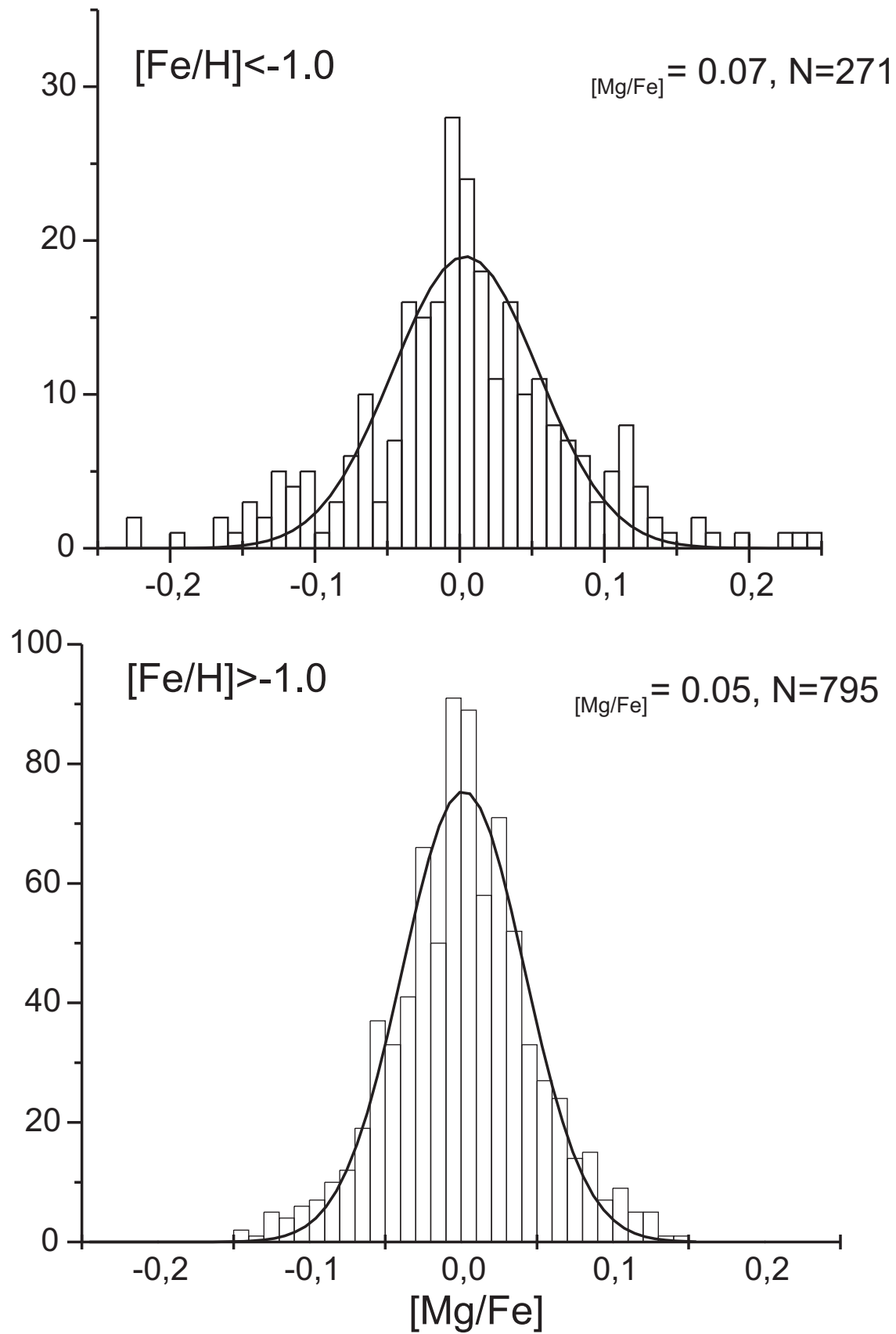}
\caption{The distributions of the $\delta$[Mg/Fe] deviations for the two
metallicity ranges. The notation is as in Fig.~2.
\hfill}
\label{fig5}
\end{figure*}

\begin{figure*}[t!]
\centering
\includegraphics[angle=0,width=0.9\textwidth,clip]{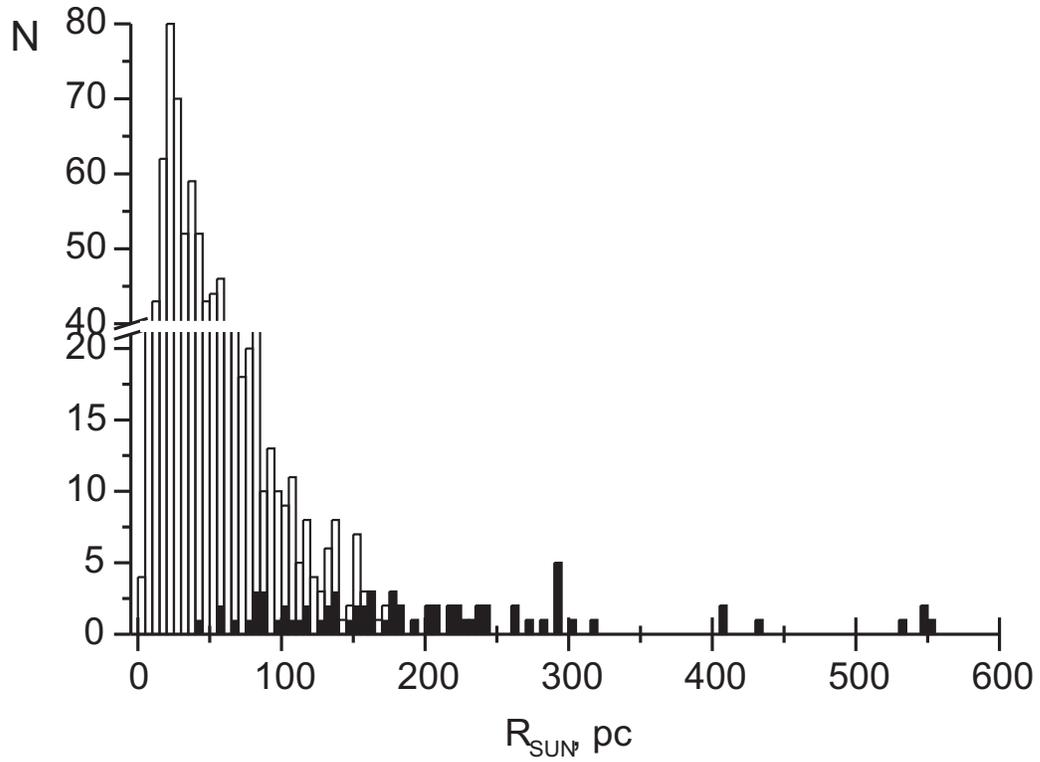}
\caption{The distribution of the distance of the catalog stars from the Sun derived from 
trigonometric parallaxes (open bars) and from photometric measurements
(filled bars).
\hfill}
\label{fig6}
\end{figure*}

\begin{figure*}[t!]
\centering
\includegraphics[angle=0,width=0.9\textwidth,clip]{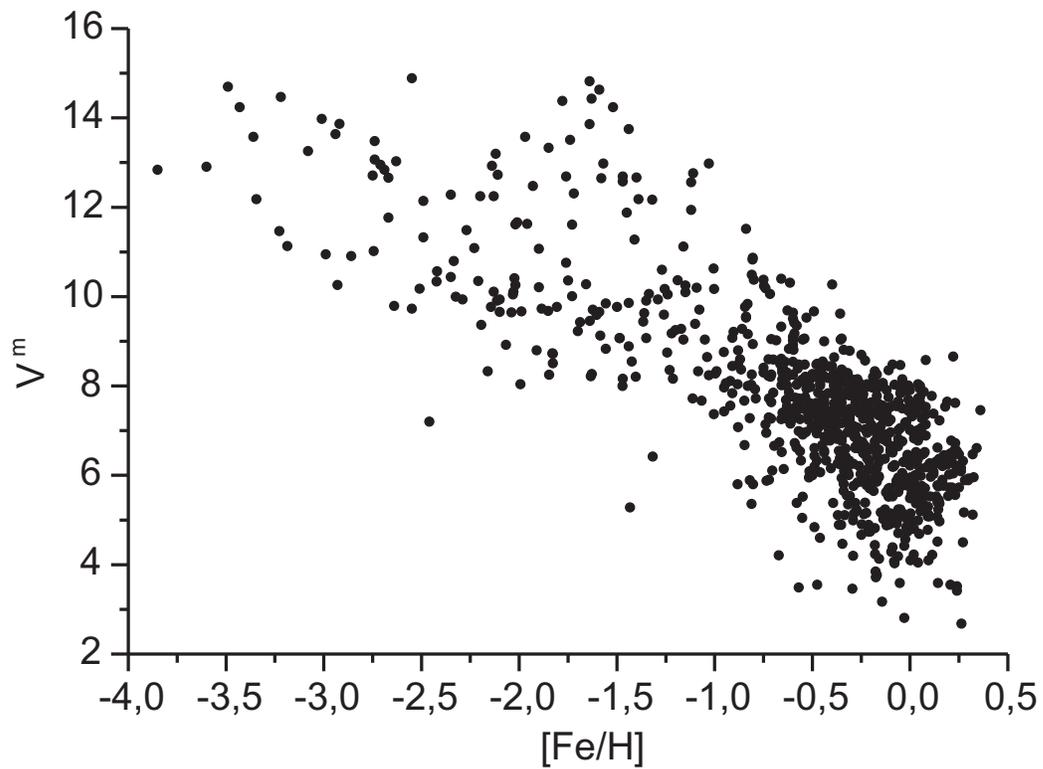}
\caption{The metallicity versus apparent magnitude.
\hfill}
\label{fig7}
\end{figure*}

\begin{figure*}
\centering
\includegraphics[angle=0,width=0.9\textwidth,clip]{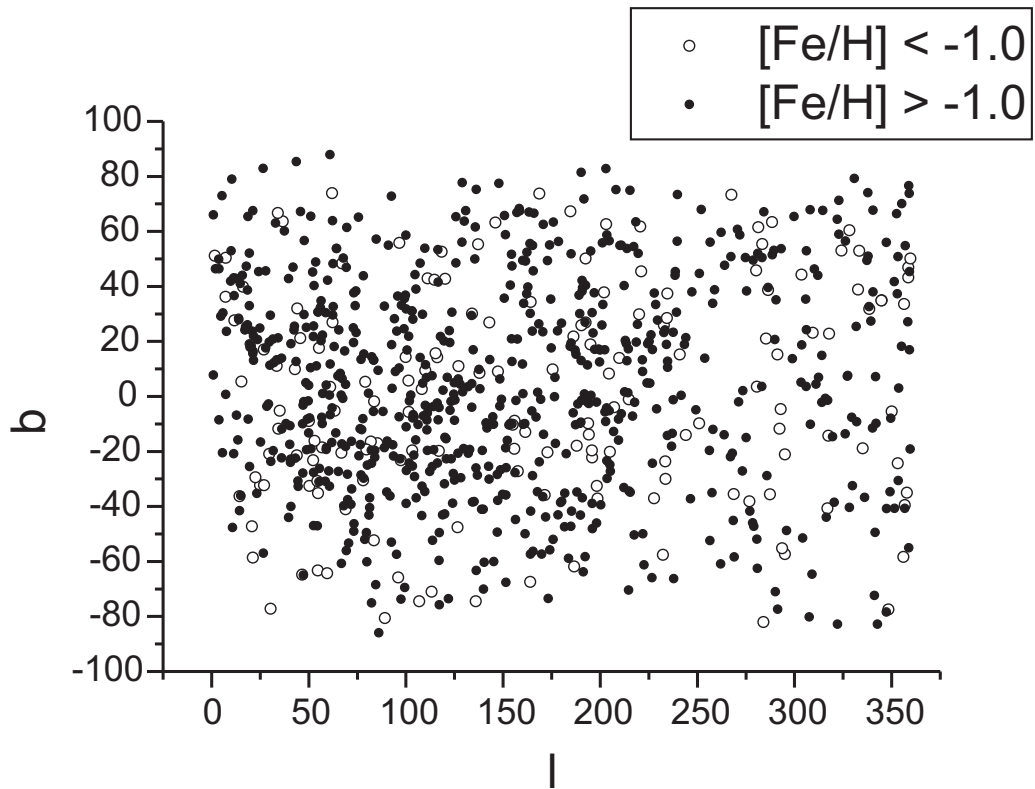}
\caption{Positions of metal-rich (filled circles) and metal-poor (open
circles) stars on the sky.
\hfill}
\label{fig8}
\end{figure*}

\begin{figure*}
\centering
\includegraphics[angle=0,width=0.9\textwidth,clip]{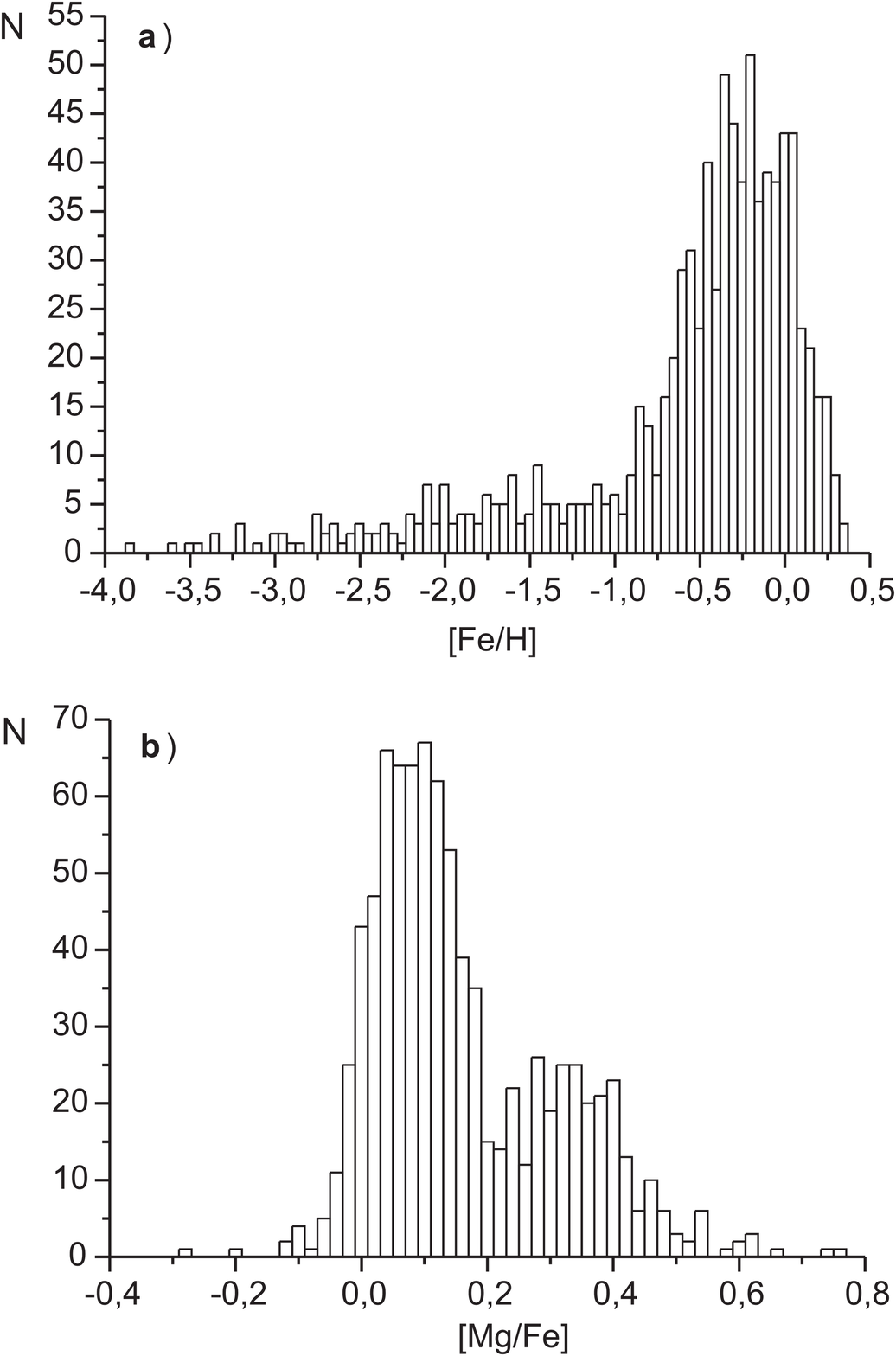}
\caption{The distributions of the (a) iron abundance and (b) relative magnesium abundances of
the catalog stars.
\hfill}
\label{fig9}
\end{figure*}

\begin{figure*}
\centering
\includegraphics[angle=0,width=0.9\textwidth,clip]{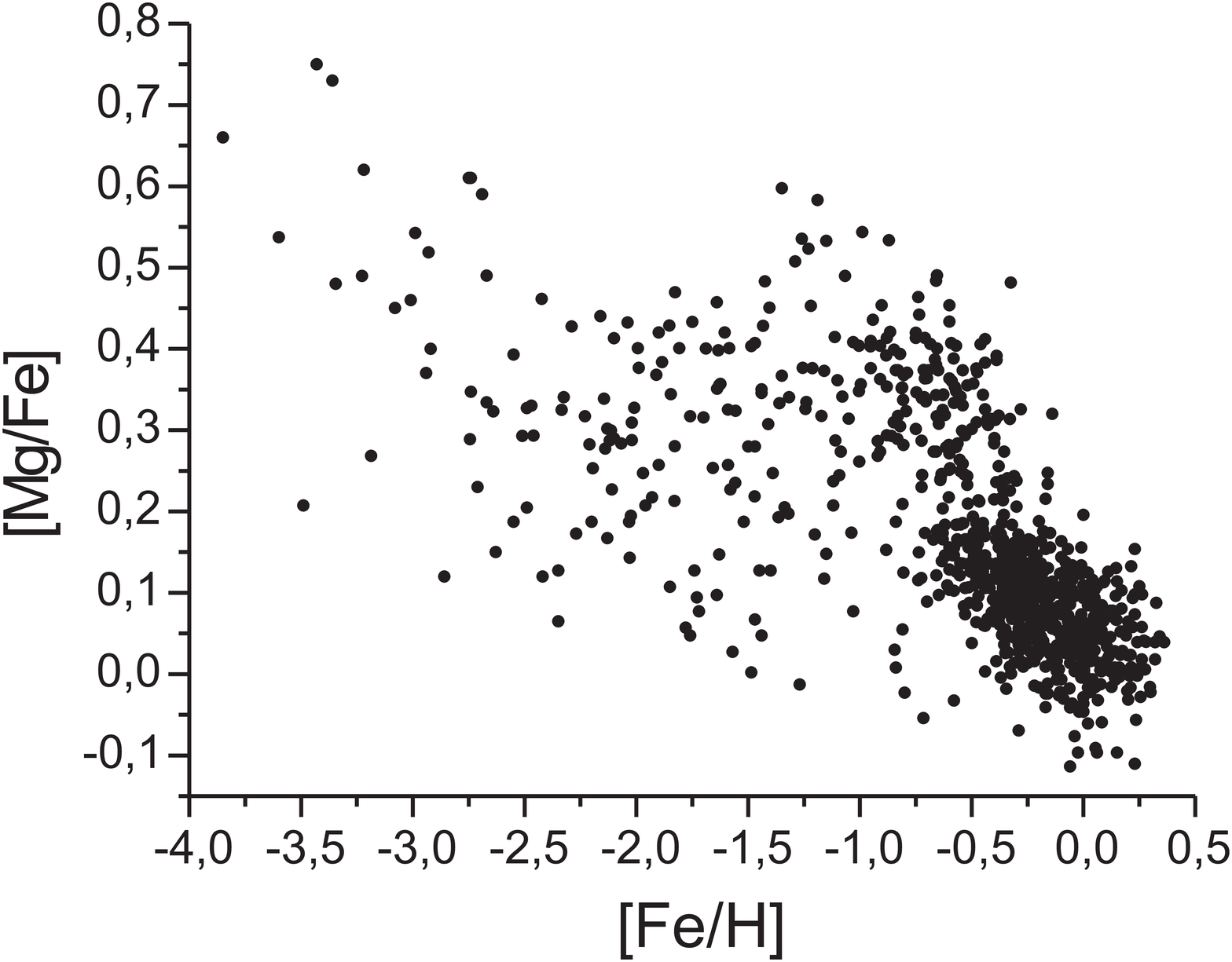}
\caption{The metallcity versus magnesium abundance for the catalog stars
stars.
\hfill}
\label{fig10}
\end{figure*}


\newpage

\begin{thebibliography}{}

\bibitem [1.]{borkova_n}
T.~V.~Borkova, V.~A.~Marsakov, Astronomy Letter, \textbf{30}, 148 (2004)

\bibitem [2.]{borkova_n}
R.~G.~Gratton, E.~Carretta, F.~Matteuchi, and C.~Sneden, in
Formation of the Galactic Halo Inside and Out, ASP Conf. Ser.,
(eds H.~Morrison and A.~Sarajedini, 1996), \textbf{92}, p.~307.

\bibitem [3.]{borkova_n}
R.~G.~Gratton, E.~Carretta, F.~Matteucci, and C.~Sneden, Astron.
and Astrophys. \textbf{358}, 671 (2000).

\bibitem [4]{Borkova_n}
K.~Fuhrmann, Astron. and Astrophys. \textbf{338}, 161 (1998).

\bibitem [5]{Borkova_n}
K.~Fuhrmann, 2000, preprint Nearby stars of the Galactic disk and
halo. II.~Munich.

\bibitem [6]{Borkova_n}
K.~Fuhrmann, New Astron. \textbf{7}, 161 (2002).

\bibitem [7]{Borkova_n}
R.~G.~Gratton, E.~Carretta, S.~Desidera, \emph{et al.}, Astron.
and Astrophys. \textbf{406}, 131 (2003).

\bibitem[8]{Borkova_n}
J.~Prochaska, S.~O.~Naumov, B.~W.~Carney, \emph{et al.}, Astron.
J. \textbf{120}, 2513 (2000).

\bibitem[9]{Borkova_n}
P.~E.~Nissen and W.~J.~Schuster, Astron. and Astrophys.
\textbf{326}, 751 (1997).

\bibitem[10]{Borkova_n}
M.~Shetrone, K.~A.~Venn, E.~Tolstou, \emph{et al.}, Astron. J.
\textbf{125}, 684 (2003).

\bibitem[11]{Borkova_n}
E.~Tolstou, K.~A.~Venn, M.~Shetrone, \emph{et al.}, Astron. J.
\textbf{125}, 707 (2003).

\bibitem[12]{Borkova_n}
N.~N.~Shimanskaya, L.~I.~Mashonkina, N.~A.~Sakhibullin Astron. Rep.
\textbf{44}, 530 (2000)

\bibitem[13]{Borkova_n}
B.~Edvardsson, J.~Andersen, B.~Gustafsson, \emph{et al.}, Astron.
and Astrophys. \textbf{275}, 101 (1993).

\bibitem[14]{Borkova_n}
N.~N.~Shimanskaya, L.~I.~Mashonkina, Astron. Rep. \textbf{45}, 100 
(2001)

\bibitem[15]{Borkova_n}
L.~Mashonkina, T.~Gehren, C.~Travaglio, and T.~Borkova, Astron.
and Astrophys. \textbf{397}, 275 (2003).

\bibitem[16]{Borkova_n}
K.~Fuhrmann, M.~Axer, and T.~Gehren, Astron. and Astrophys.
\textbf{301}, 492 (1995).

\bibitem[17]{Borkova_n}
Y.~Q.~Chen, P.~E.~Nissen, G.~Zhao, \emph{et al.}, Astron. and
Astrophys. Suppl. Ser. \textbf{141}, 491 (2000).

\bibitem[18]{Borkova_n}
Y.~Q.~Chen, G.~Zhao, P.~E.~Nissen, \emph{et al.}, Astron. and
Astrophys. \textbf{591}, 925 (2003).

\bibitem[19]{Borkova_n}
T.~V.~Mishenina and V.~V.~Kovtyukh, Astron. and Astrophys.
\textbf{370}, 951 (2001).

\bibitem[20]{Borkova_n}
T.~V.~Mishenina, '.~Soubiran, V.~V.~Kovtyukh, S.~A.~Korotin,
Astron. and Astrophys. \textbf{418}, 551 (2004).

\bibitem[21]{Borkova_n}
G.~Clementini, R.~Gratton, E.~Carretta, and C.~Sneden, Monthly
Not. Astron. Soc. \textbf{302}, 22 (1999).

\bibitem[22]{Borkova_n}
E.~Carretta, R.~Gratton, and C.~Sneden, Astron. and Astrophys.
\textbf{356}, 238 (2000).

\bibitem[23]{Borkova_n}
P.~E.~Nissen, B.~Gustafsson, B.~Edvardsson, and G.~Gilmore,
Astron. and Astrophys. \textbf{440} (1994).

\bibitem[24]{Borkova_n}
R.~Ottmann, M.~J.~Pfeiffer, and T.~Gehren, Astron. and Astrophys.
\textbf{338}, 661 (1998).

\bibitem[24]{Borkova_n}
G.~Zhao and T.~Gehren, Astron. and Astrophys. \textbf{362}, 1077
(2000).

\bibitem[26]{Borkova_n}
T.~Gehren, Y.~C.~Liang, J.~R.~Shi, \emph{et al.}, Astron. and Astrophys.
\textbf{413}, 1045 (2004).

\bibitem[27]{Borkova_n}
P.~Magain, Astron. and Astrophys. \textbf{209}, 211 (1989).

\bibitem[28]{Borkova_n}
E.~Jehin, P.~Magain, C.~Neuforge, \emph{et.al.}, Astron. and
Astrophys. \textbf{341}, 241 (1999).

\bibitem[29]{Borkova_n}
A.~Stephens, Astron. J. \textbf{117}, 1771 (1999).

\bibitem[30]{Borkova_n}
A.~Stephens and A.~M.~Boesgaard, Astron. J. \textbf{123}, 1647
(2002).

\bibitem[31]{Borkova_n}
B.~W.~Carney, J.~S.~Wright, C.~Sneden, \emph{et al.}, Astron.
J. \textbf{114}, 363 (1997).

\bibitem[32]{Borkova_n}
T.~Idiart and F.~Thevenin, Astrophys. J. \textbf{541}, 207 (2000).

\bibitem[33]{Borkova_n}
T.~Bensby, S.~Feltzing, and I.~Lundstrom, Astron. and Astrophys.
\textbf{410}, 527 (2003).

\bibitem[34]{Borkova_n}
S.~V.~Ermakov, Ph. dissetation in Mathematical Fhysics (SAO, 2002)

\bibitem[35]{Borkova_n}
S.~G.~Ryan, J.~E.~Norris, and M.~S.~Bessell, Astron. J.
\textbf{102}, 303 (1991).

\bibitem[36]{Borkova_n}
S.~G.~Ryan, J.~E.~Norris, and T.~C.~Beers, Astrophys. J.
\textbf{471}, 254 (1996).

\bibitem[37]{Borkova_n}
R.~Gratton, E.~Carretta, R.~Claudi, \emph{et al.}, Astron. and
Astrophys. \textbf{404}, 187 (2003).

\bibitem[38]{Borkova_n}
B.~E.~Reddy, J.~Tomkin, L.~Lambert, \emph{et al.}, Monthly Not.
Astron. Soc. \textbf{340}, 304 (2003).

\bibitem[39]{Borkova_n}
B.~Hauck and M.~Mermilliod, Astron. and Astrophys. Suppl. Ser.
\textbf{129}, 431 (1998).

\bibitem[40]{Borkova_n}
S.~Castro, R.~M.~Rich, M.~Grenon, \emph{et. al.}, Astron. J.
\textbf{114}, 376 (1997).

\bibitem[41]{Borkova_n}
The Hipparcos and Tycho Catalogues, ESO, 1997.

\bibitem[42]{Borkova_n}
N.~V.~Kharchenko, All-sky Compiled Catalogue of 2.5 million stars,
KFNT \textbf{17}, 409 (2001).

\bibitem[43]{Borkova_n}
J.~R.~Myers, C.~B.~Sande, A.~C.~Miller, Jr. W.~H.~Warren, and
D.~A.~Tracewell, 2002, Sky2000 Catalogue, version 4.

\bibitem[44]{Borkova_n}
B.~W.~Carney, D.~W.~Latham, J.~B.~Laird, and L.~A.~Aguilar,
Astron. J. \textbf{107}, 2240 (1994).

\bibitem[45]{Borkova_n}
B.~Nordstrom, M.~Mayor, J.~Andersen, \emph{et al.}, Astron. and
Astrophys. \textbf{418}, 989 (2004).

\bibitem[46]{Borkova_n}
T.~C.~Beer, M.~Chiba, Y.~Yoshi, \emph{et al.}, Astron. J.
\textbf{119}, 2866 (2000).

\bibitem[47]{Borkova_n}
G.~A.~Bakos, K.~C.~Sahu, and P.~Nemeth, Astrophys. J.~Suppl. Ser.
\textbf{141}, 187 (2002).

\bibitem[48]{Borkova_n}
D.~L.~Nidever, G.~W.~Marcy, R.~P.~Butler, D.~A.~Fischer, and
S.~S.~Vogt, Astrophys. J.~Suppl. Ser. \textbf{141}, 503 (2002).

\bibitem[49]{Borkova_n}
M.~Barbier-Brossat and P.~Figon, Astron. and Astrophys.
\textbf{142}, 217 (2000).

\bibitem[50]{Borkova_n}
M.~Barberi and R.~G.~Gratton, Astron. and Astrophys. \textbf{384},
879 (2002).

\bibitem[51]{Borkova_n}
K.~U.~Ratnatunga, J.~N.~Bahcall, and S.~Casrtano, Astrophys. J.
\textbf{291}, 260 (1989).

\bibitem[52]{Borkova_n}
C.~Allen and A.~Santillan, Rev. Mex. Astron. y Astrofis.
\textbf{22}, 255 (1991).

\end{thebibliography}
\end{document}